# Origin of the Superconductivity in the Y-Sr-Ru-O and Y-Sr-Cu-O Systems

E.Galstyan[a], Y.Y.Xue[a], M.N.Iliev[a], Y.Sun[a], and C.W.Chu[a,b,c]

[a] Department of Physics and Texas Center for Superconductivity, University of Houston, Houston, TX 77204-5002, USA
[b] Lawrence Berkeley National Laboratory, 1 Cyclotron Road, Berkeley, CA 94720, USA
[c]Hong Kong University of Science and Technology, Hong Kong, China

## Abstract

We report on the structural, magnetic, and Raman-scattering studies of double perovskite structure $Sr_2Y(Ru_{1-x}Cu_x)O_{6-\delta}$ systems made by systematic synthesis processes with various numbers of doping concentrations and sintering temperatures. We observed different behaviors resulting from the different thermal treatments. In particular, superconductivity in Cu-doped $Sr_2YRuO_6$ has been observed only for partially melted ceramic materials. We show that superconductivity is associated with the 1:2:3 phase ($YSr_2Cu_3O_t$), similar to that of Y-Sr-Cu-O samples sintered at high temperature.

## I. Introduction

The nature and origin of the superconductivity (SC) in perovskite-based cuprates have attracted much interest among researchers in recent years. The common features of the high-temperature superconducting cuprates are sequences of $CuO_2$ layers that upon sufficient doping become superconducting. It has been one major challenge to search for the possibility of finding new superconductors that contains no Cu-O planes. Initially, this interest was caused by the discovery of exotic SC in $Sr_2RuO_4$ [1]. Later, it was found that other ruthenates have very interesting magnetic and electric properties. Among them there has been particular interest in the Ru-based double perovskite structure $Sr_2YRuO_6$ systems with minor Cu-doping on Ru ion sites [2-5]. The parent insulator compounds $Sr_2YRuO_6$ can be obtained from the itinerant ferromagnet $SrRuO_3$ by replacing each second Ru ion with a nonmagnetic Y ion [6-9]. The antiferromagnetic (AFM) transition temperature ($T_N$) is 26 K and there is a spin-flop like transition below $T_M$~17 K [8]. An effective paramagnetic moment $\mu_{eff}$ was found to be 3.13 $\mu_B/Ru^{5+}$, which was attributed a spin-orbit coupling [7]. Below $T_N$=26 K the face-centered cubic (FCC) lattice of $Ru^{5+}$ spins experiences ordering to produce a Type-I AFM structure with a weak ferromagnetic component [8, 9]. In this structure, the (001) ferromagnetic (FM) planes exhibit antiparallel ordering along the *c* axis.

Intriguingly, SC has been reported in $Sr_2Y(Ru_{1-x}Cu_x)O_{6-\delta}$ ceramic materials with an onset transition temperature of $T_C$~ 45-49 K for x=0.03-0.50 [2-5]. Addressing the question of the nature of SC, J. D. Dow and D. R. Harshman [5, 10, 11] proposed a model in which superconducting condensation occurs in the SrO planes with the $(Ru_{1-x}Cu_x)O$ plans magnetically ordered. While the model appears to be in disagreement with many research works [12-16], the exact reason for SC in Y-Sr-Ru-O system is still uncertain.

In order to gain additional insight into the superconducting properties of Cu-doped $Sr_2YRuO_6$ ceramic compounds, we present the results of magnetic measurements and Raman-scattering investigation. The crystal structure, lattice parameters, and detailed microstructure were analyzed by X-ray powder diffraction method (XRD) and scanning electron microscopy (SEM). In this work we provide experimental evidence that the SC behavior of these materials is due to a minor impurity phase with the stoichiometry of $YSr_2Cu_3O_t$ (YSCO). We show that both superconductivity and YSCO grains appear only for synthesis temperatures higher than a local-melting temperature identifiable by the differential thermal analyzer (DTA) trace. In addition, the YSCO grains, which are thermodynamically unstable under ambient pressure, appear within dense surroundings, where the melting indications are obvious. The stresses during the solidification seem to play roles. These may also explain the earlier reports of trace SC in Y-Sr-Cu-O samples synthesized at ambient pressure [17, 18].

## II. Experimental details

Ceramic samples with nominal compositions $Sr_2Y(Ru_{1-x}Cu_x)O_{6-\delta}$ (x=0.0, 0.1, 0.5) and Y-Sr-Cu-O ( $YSrCuO_t$, $YSr_2Cu_2O_t$, $Y_{1.5}Sr_2Cu_2O_t$ ) were prepared by a solid-state reaction technique. Prescribed amounts of $Y_2O_3$, $SrCO_2$, $RuO_2$, and CuO were mixed, pressed into pellets, and preheated at 1000 °C for 1 day (850 °C for Y-Sr-Cu-O). The first and second batches of $Sr_2Y(Ru_{1-x}Cu_x)O_{6-\delta}$ were cooled, reground, and sintered under an oxygen atmosphere for 1 day at 1290 °C and 1360 °C, respectively, and finally the samples were slowly cooled at 15 °C per hour to room temperature. Some of the Y-Sr-Cu-O calcinated powders were pelletized and heated at 990 °C for 3 days under ambient pressure in an oxygen atmosphere. Other Y-Sr-Cu-O calcinated materials were reheated at 1130 °C for 6 hours in flowing oxygen.

X-ray powder diffraction (XRD) was used to identify the bulk phase, approximate composition, and lattice parameters. Diffraction patterns were obtained with CuKα radiation (wavelength $\lambda$=0.154178 nm) over a range of 15°< $2\theta$< 80° using 0.04° steps. The microstructure and the phase integrity of the materials were investigated by scanning electron microscopy (SEM) and by a Genesis energy dispersive x-ray analysis (EDAX) device attached to the SEM. The melting temperature was determined with a differential thermal analyzer (DTA). Magnetic susceptibility data were obtained with a superconducting quantum interference device (SQUID) magnetometer. Polarized Raman spectra were collected under microscope (focus spot size 1-3 μm, $\lambda_{exc}$= 488 nm) from the polished surface of the bulk materials at room temperature.

## III Results and Discussion

### (i) $Sr_2Y(Ru_{1-x}Cu_x)O_{6-\delta}$

Figure 1 shows the x-ray diffraction patterns of $Sr_2Y(Ru_{1-x}Cu_x)O_{6-\delta}$ for x=0, 0.1, and 0.5. The major reflections can be indexed with orthorhombic symmetry. Within the instrumental resolution of a few percent, the absence of impurity lines for x=0 and 0.1 shows high phase purity and suggests the existence of Ru-Cu solid solution in this double-perovskite component. The presence of impurity peaks at x=0.5 (marked with asterisks), on the other hand, indicates a Cu solubility-limit below 0.5, although it was impossible to identify the corresponding phases due to the low intensity of diffraction data. Comparing the peak intensities, we roughly estimate that the secondary phases occupy less than 20% of the volume. Least-square fits to the *Pbnm* space group shows a systematic decrease of the orthorhombic distortion and an increase of the lattice parameter $c$ with the Cu content, *i.e.* $a:b:c$ = 5.712(2):5.773(4):8.164(9) Å; 5.715(8):5.771(8):8.165(6) Å; and 5.723(1):5.764(6):8.192(9) Å for x = 0, 0.1, and 0.5, respectively. We have noticed slight discrepancies in lattice constants for x=0 in comparison with cell parameters $a$=5.7690(6) Å, $b$=5.7777(6) Å, and $c$=8.1592(9) Å published in Ref. 8. Similar discrepancies and trends in lattice constant with Cu content have been reported in a previous publication [2].

Figure 2 shows the result of DTA analyses of multiphase $Sr_2Y(Ru_{0.5}Cu_{0.5})O_{6-\delta}$ compound (powder) with increasing temperature in air. Endothermic peaks appear near 1130 and 1275 °C. The Cu-rich sample sintered at 1360 °C became harder and looked partially melted at the bottom.

The temperature dependence of the dc magnetization of $Sr_2Y(Ru_{1-x}Cu_x)O_{6-\delta}$ (x=0.1 and x=0.5) samples measured at 7 Oe, and the real ac susceptibility curve (at $H_{dc}$=0 Oe, $H_{ac}$=3 Oe) of $Sr_2Y(Ru_{0.5}Cu_{0.5})O_{6-\delta}$ are presented in Fig. 3.

The general behavior of the dc magnetic susceptibility curves (Fig.3a) of the $Sr_2Y(Ru_{0.9}Cu_{0.1})O_{6-\delta}$ sample is very similar to that of the parent compound $Sr_2YRuCuO_6$ [8, 9]. The system orders AFM at $T_N$=25-30 K and yet shows conspicuously weak ferromagnetism by large irreversibility. For x=0.5, a similar magnetic bump at 25 K is observed. The ZFC and FC curves, however, split at higher temperature (~ 58 K; see the inset in Fig.3a). These results strongly suggest that the magnetic structure remains unchanged for Cu-doped samples and it is consistent with data published in the literature [2, 5, 11].

In all the samples tested, no trace superconductivity has been observed for x = 0.1, in rough agreement with previous reports [2, 5, 11]. The situation is different in the case of x = 0.5. The negative signals in the ZFC curve at 5 K, which correspond to a screened volume-fraction of 4% or larger, clearly indicate a SC component (Fig.3a), which is consistent with both the negative FC magnetization around 40 K and the real part, $\chi$', of the ac susceptibility (Fig.3b). It is interesting to note that the onset of the SC occurs at 64 K, as indicated by the downturns in both the FC magnetization (flux-expulsion) and the $\chi$'.

The magnetization $M(H)$ plots at 5 K of the compounds with x=0.1 and 0.5 up to 3500 Oe and 5 Tesla (T) are shown respectively in Fig.4 and its inset. The virgin curve for the x=0.5 sample synthesized at 1360 °C clearly shows a SC signature. The diamagnetic signal starts decreasing above applied fields of 280 Oe, and turns to zero at 1800 Oe. The compound seems to have a lower critical field ($Hc_1$) of around 280 Oe. The shielding fraction (SF) deduced from the isothermal $M(H)$ curve is ~ 1.5% of the $-1/4\pi$ value, indicating that only a small fraction of the material becomes superconducting. For the samples synthesized at 1290 °C (not shown in the figure), on the other hand, no indications of superconductivity have been noticed. Both a high doping-level, *i.e.* above the solubility limit of Cu, and a high synthesis temperature seem to be necessary for the appearance of superconductivity.

To explore the origin of this minor superconductivity, the grain morphology of the polished $Sr_2Y(Ru_{0.5}Cu_{0.5})O_{6-\delta}$ sample, which shows the clear superconducting signal, was determined by SEM .

Two types of granular structures can be clearly identified, although both have a similar grain size of 10-13 μm. One type (marked as I in Fig. 5) consists of more or less round grains with many voids along the grain boundaries. This is rather similar to those of the samples synthesized at lower temperature, which is also typically characteristic of ceramics below the melting temperature. Another type (marked as II), however, consists of thick wavering sheets with well defined inter-grain boundaries characterized by sharp edges. We interpret these as evidence of partial melting during synthesis. It is worth noting that there are practically no voids along the type II grain boundaries.

EDAX analysis further reveals a distinct difference between the two types. For the type I grains, no systematic intra-grain stoichiometry-variation is detected although the ratio Sr:Y:Ru:Cu does vary from one grain to another: a normalized ratio of 2:1.2:1.3:0.4 represents most of the grains. Some spherical grains with a poor Cu concentration also have been observed. The typical normalized stoichiometric ratio of these particles is 2:1.1:1.4:0.01, which is similar to the structure of the parent $Sr_2YRuO_6$ compound. In addition to the major phase in $Sr_2Y(Ru_{0.5}Cu_{0.5})O_{6-\delta}$, we found that the type II grains correspond to a secondary phase with an general stoichiometric relation of 1.3:2:3.7:0.2 (Y:Sr:Cu:Ru), which is close to the 1:2:3 phase and consistent with $YSr_2Cu_3O_t$ (YSCO). Apparently, this secondary phase for x=0.5 is below 5% and therefore has not been identified by XRD measurements. Detailed EDAX analysis of the Cu-rich type II grain shows that deficiency in Ru content continuously increases with distance from the center of the grains, and almost pure YSCO structure [1.2(Y):2(Sr):3.3(Cu):0.03(Ru)] is observed on the boundaries. Both the Ru depletion and redistribution occur in the melted region, while no visible deviation in Y and Sr composition appears. The observed Ru-depletion may be due to sublimation of ruthenium oxide in the oxidizing atmosphere [19], and the Ru/Cu redistribution will be a natural result of the correlation between the melting temperature and Cu-doping. Thus, partially melted grains may lead to the Ru-absence and, therefore, to the YSCO phase on the boundaries.

The polycrystalline sample $Sr_2Y(Ru_{0.5}Cu_{0.5})O_{6-\delta}$ was studied by Raman spectroscopy. Further verification of the presence of a YSCO minority phase comes from the Raman spectra probed at type I and type II grains. As illustrated in Fig. 6, the spectra of the two types of grains differ significantly. While those from type I are in good agreement with the spectra of polycrystalline $Sr_2Y(Ru_{0.9}Cu_{0.1})O_{6-\delta}$, published in Ref. [20], the spectra from the type II grains are consistent with those reported earlier for YSCO [21].

### (ii) The Y-Sr-Cu-O system

In an attempt to understand the appearance of YSCO under ambient pressure, we investigated the Y-Sr-Cu-O system, particularly the samples with nominal compositions $YSrCuO_t$, $YSr_2Cu_2O_t$, and $Y_{1.5}Sr_2Cu_2O_t$. Superconductivity was used as the experimental indicator of the YSCO phase. In general, it is believed that pure YSCO without chemical doping can be synthesized only under high pressure in the temperature range 1050-1130 °C [22-24]. On the other hand, trace SC has been observed in a multiphase sample with nominal composition $YSrCuO_t$ synthesized under ambient pressure above 1200 °C [18, 25]. In the absence of detailed structural analysis, the authors have tentatively assigned the 80 K SC phase to the YSCO and AFM transition to $Y_2Cu_2O_5$ magnetic phase around 14 K. Our magnetization data (Fig. 7) of multiphase Y-Sr-Cu-O compounds closely resembles the earlier reported results [18, 25]. Namely, all Y-Sr-Cu-O materials sintered at 1130 °C show similar onsets of SC. The estimated SF fraction for all SC samples at 5 K is around 1%. For samples processed at the lower temperature of 990 °C, however, the SC phase did not form.

The grain morphology of multiphase Y-Sr-Cu-O compounds has been observed by SEM. An example of the granular structures of SC and non SC $YSrCuO_t$ samples is shown in Fig. 8. The changes in grain structures with the sintered temperature are well pronounced. One represents the coupling between partially melted grains (Fig.8a), which are closely packed without inter-grain voids, while the other type (Fig.8b) involves well-coupled grains with structurally intact boundaries. EDAX analysis, similar to that of $Sr_2Y(Ru_{0.5}Cu_{0.5})O_{6-\delta}$ samples, indicates the presence of the YSCO structure only in samples sintered at high temperature. The internal stresses during the solidification, therefore, may offer a natural interpretation for the formation of the YSCO phase.

## Conclusion

We have shown that superconductivity in $Sr_2Y(Ru_{0.5}Cu_{0.5})O_{6-\delta}$ and various Y-Sr-Cu-O systems appears only in partially melted samples under ambient pressure. The superconducting behavior is tentatively attributed to a minor $YSr_2Cu_3O_t$ phase, which occurs on grain boundaries in the first system due to the depletion and redistribution of Ru in partially melted grains, and occurs in the latter system due to the internal stress during solidification.

## Acknowledgements

The work in Houston is supported in part by the T. L. L. Temple Foundation, the John J. and Rebecca Moores Endowment, and the State of Texas through the Texas Center for Superconductivity at the University of Houston; and at Lawrence Berkeley Laboratory by the Director, Office of Science, Office of Basic Energy Sciences, Division of Materials Sciences and Engineering of the U.S. Department of Energy under Contract No. DE-AC03-76SF00098.

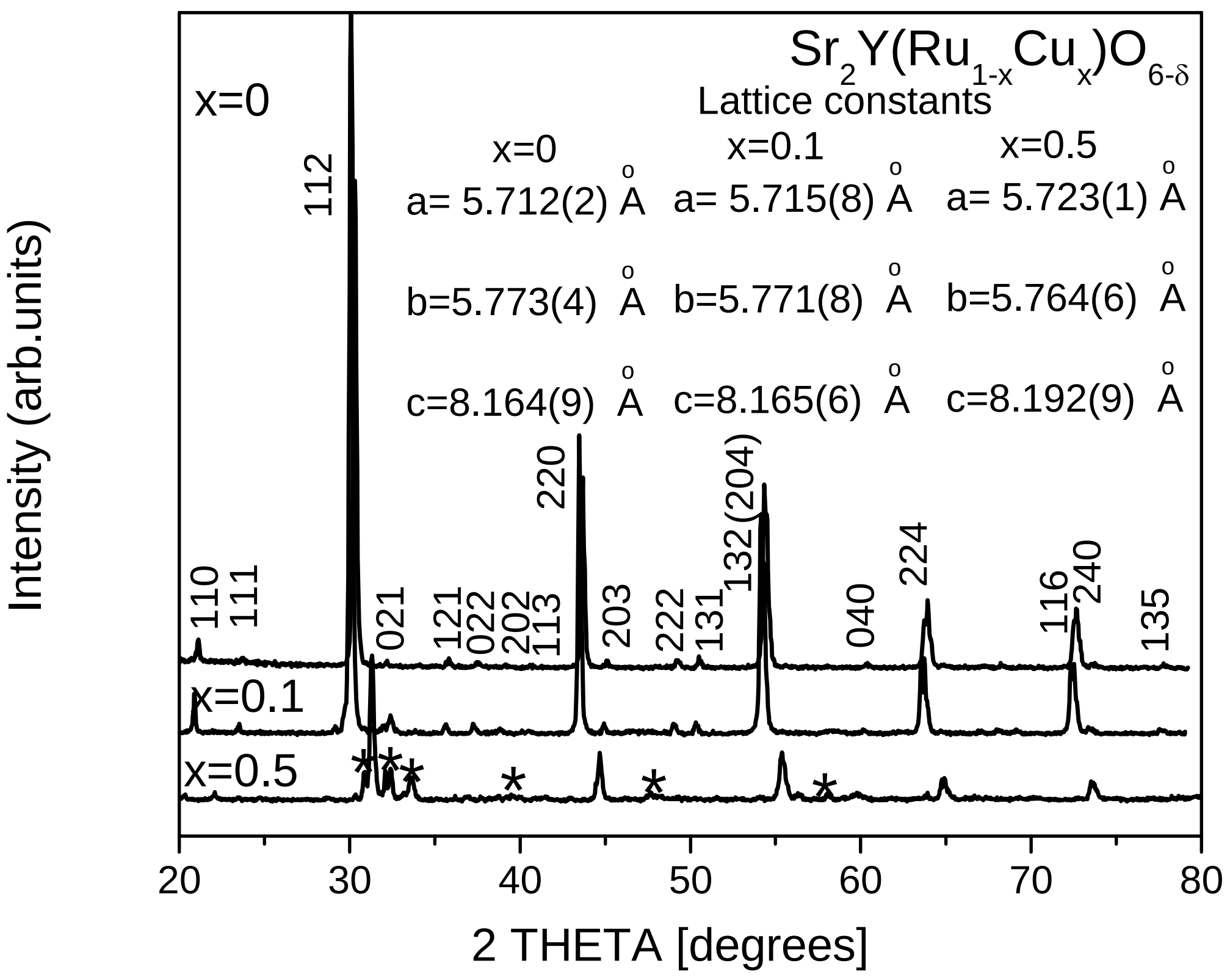

Fig.1. Powder X-ray diffraction patterns of $Sr_2Y(Ru_{1-x}Cu_x)O_{6-\delta}$ (x=0, 0.1, and 0.5). Secondary phases are marked by asterisks.

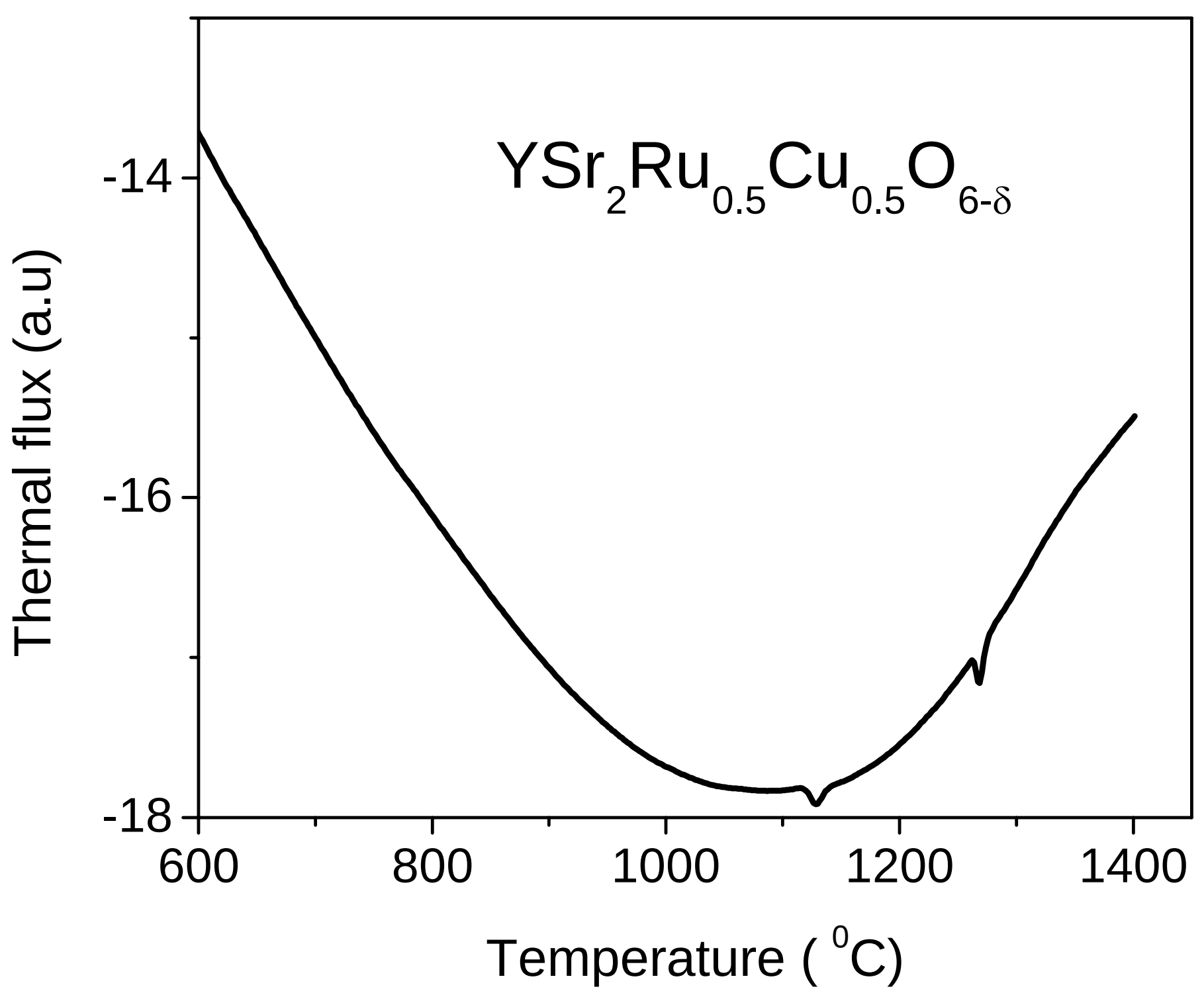

Fig.2 DTA curves for $Sr_2Y(Ru_{0.5}Cu_{0.5})O_{6-\delta}$ measured in an increasing temperature process in air.

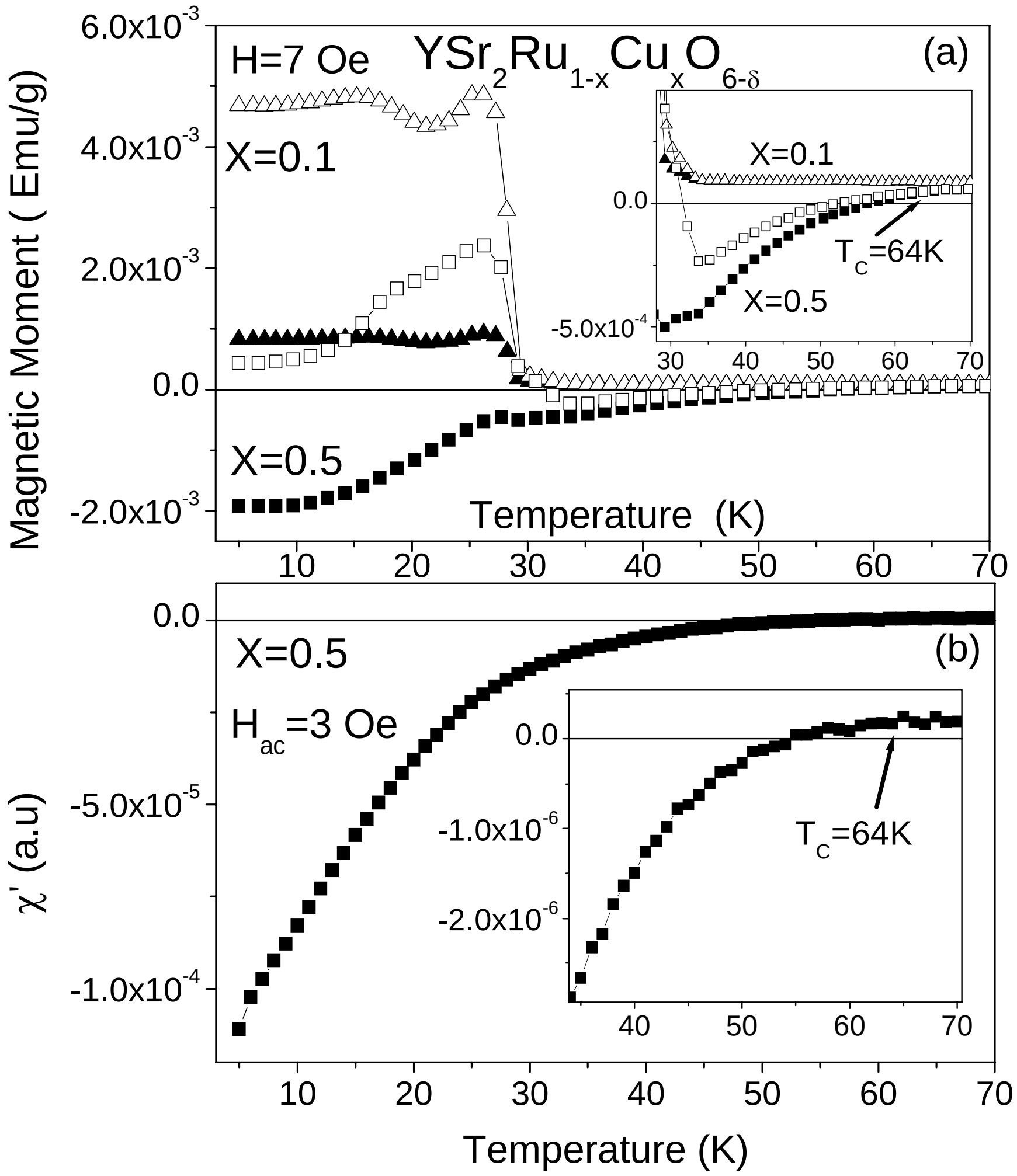

Fig.3 (a) ZFC (solid symbols) and FC (open symbols) magnetization curves for $Sr_2Y(Ru_{1-x}Cu_x)O_{6-\delta}$ with x=0.1 (triangle symbols) and x=0.5 (square symbols). Inset: ZFC and FC magnetization between 28 and 70 K. (b) The real part of the ac susceptibility (at $H_{dc}$=0 Oe, $H_{ac}$=3 Oe) of $Sr_2Y(Ru_{0.5}Cu_{0.5})O_{6-\delta}$. The inset shows the onset of superconductivity in an expanded scale.

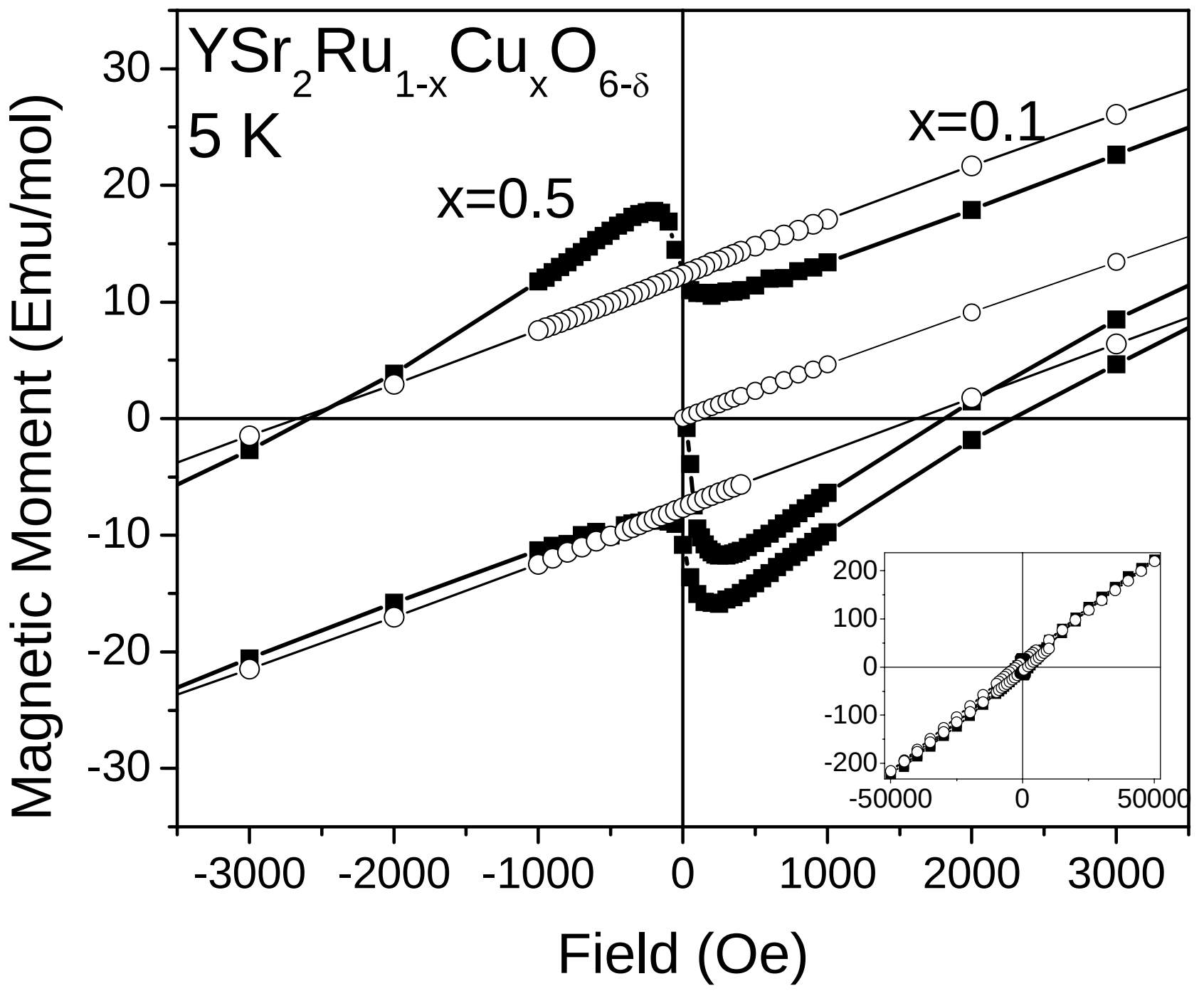

Fig.4 *M(H)* plot at 5 K for $Sr_2Y(Ru_{1-x}Cu_x)O_{6-\delta}$ with x=0.1 (open circle) and x=0.5 (solid circle); the applied fields are in the range -3500 Oe ≤ H≤3500 Oe. Inset: *M(H)* plot in the range -5 Tesla ≤ H≤ 5 Tesla.

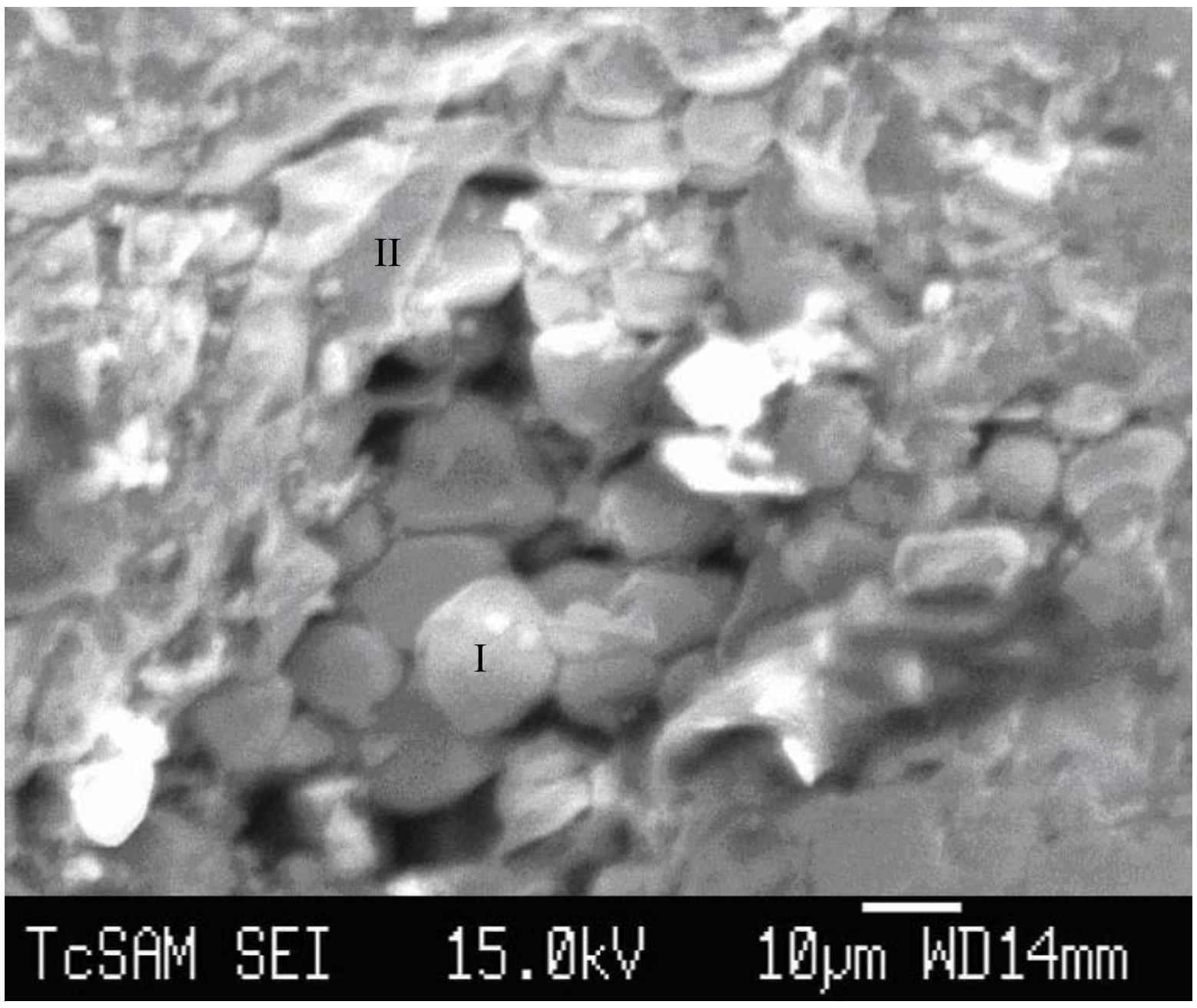

Fig.5 SEM images of $Sr_2Y(Ru_{0.5}Cu_{0.5})O_{6-\delta}$. The spherical grains with structurally intact boundaries are marked as type I. The partially melted grains without inter-grain voids are marked as type II.

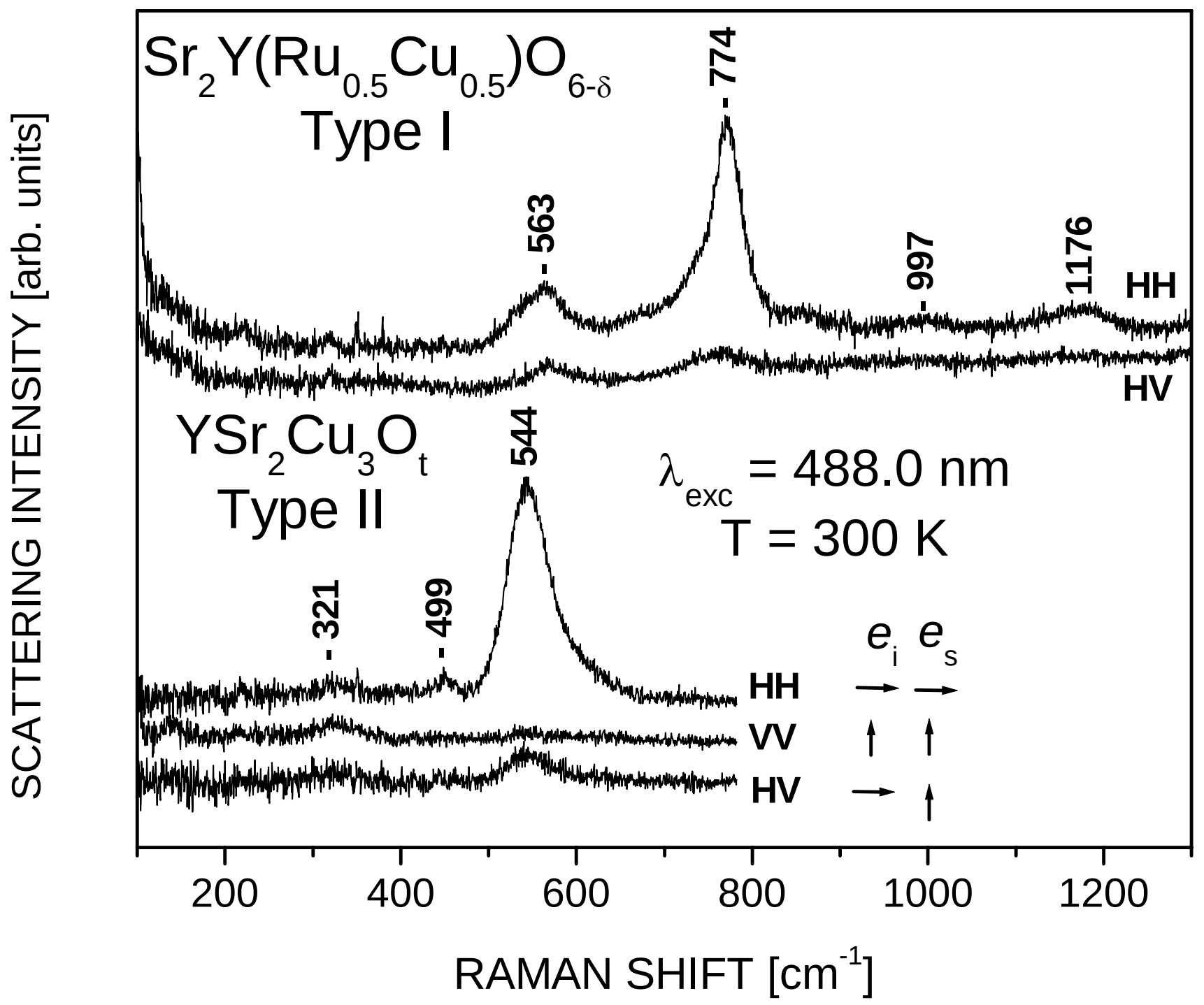

Fig.6 Polarized Raman spectra of $Sr_2Y(Ru_{0.5}Cu_{0.5})O_{6-\delta}$ (nominal composition) as obtained from type I and type II grains at room temperature.

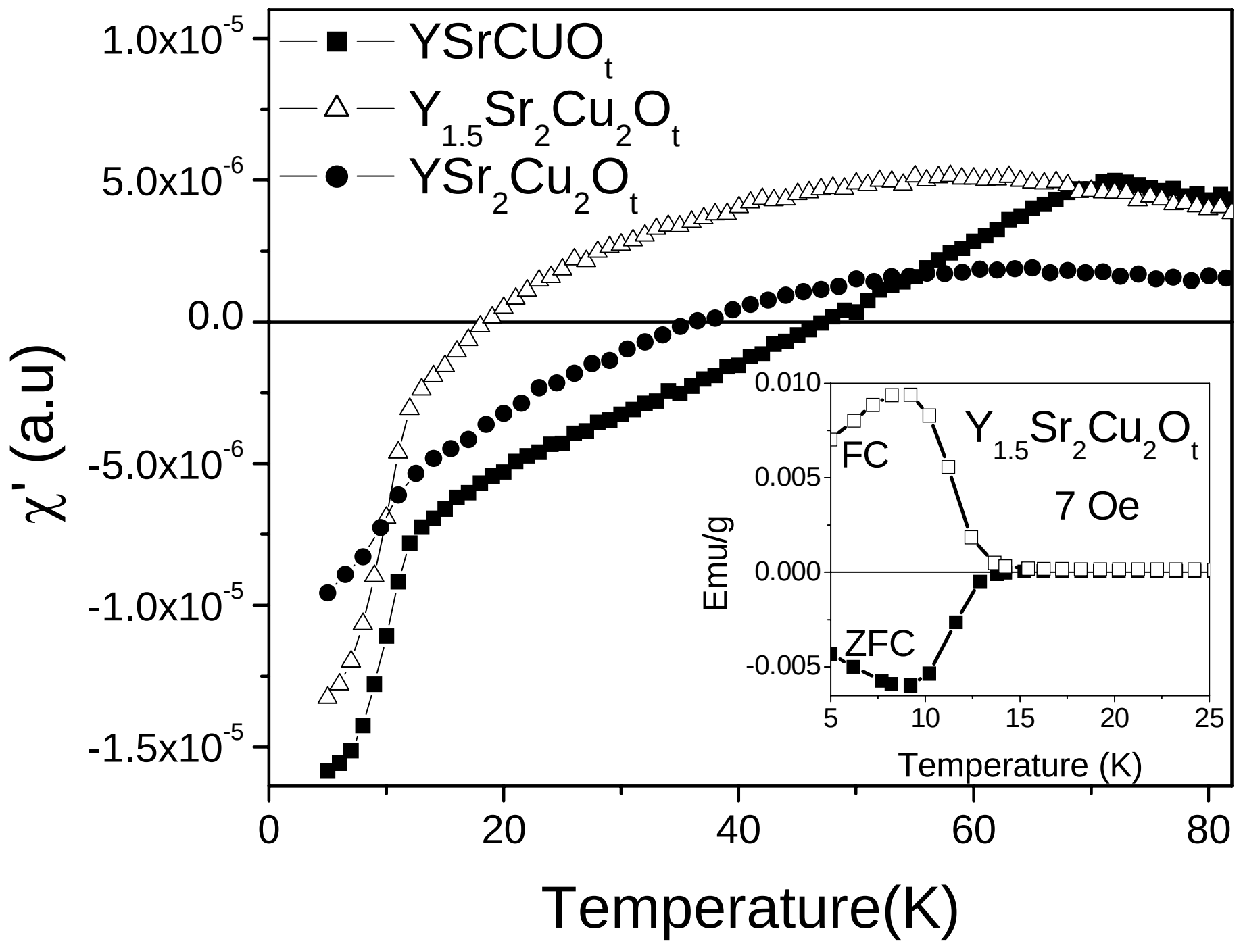

Fig.7 The real ac susceptibility curves (at $H_{dc}$=0 Oe, $H_{ac}$=3 Oe) of nominal compounds $YSrCuO_t$, $YSr_2Cu_2O_t$, and $Y_{1.5}Sr_2Cu_2O_t$. Inset: ZFC and FC magnetization curves of $Y_{1.5}Sr_2Cu_2O_t$ (nominal composition) at applied field 7 Oe .

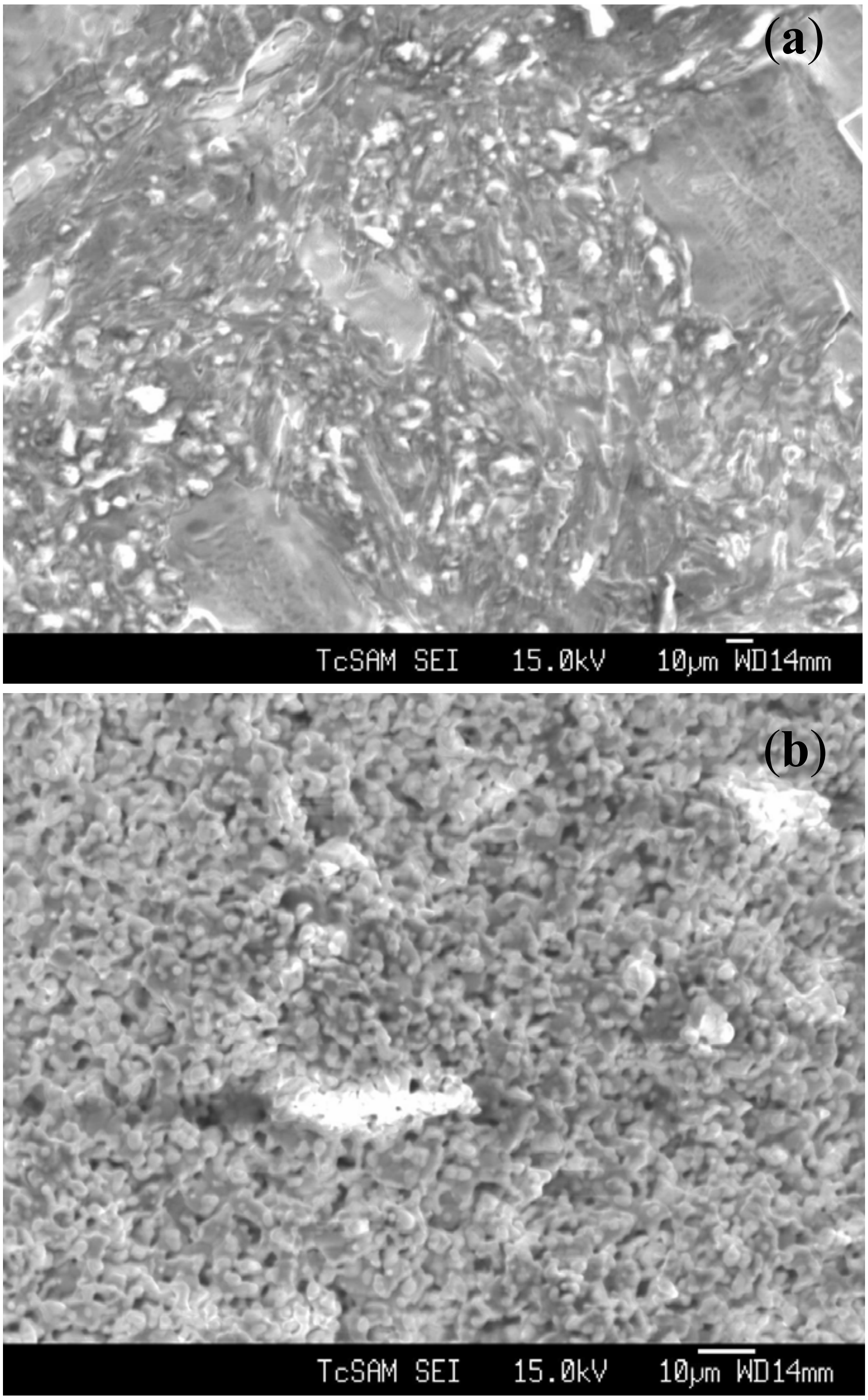

Fig. 8 SEM image of (a) SC (sintered at 1130 °C) and (b) non SC (sintered at 990 °C) $YSrCuO_t$ samples (nominal composition).